\documentclass[apsrev4-1,prd,twocolumn,showpacs]{revtex4-1}
\usepackage{graphicx}
\usepackage{dcolumn}
\usepackage{bm}
\usepackage{natbib}
\usepackage{multirow}
\newcommand{\beq}{\begin{equation}}
\newcommand{\eeq}{\end{equation}}
\newcommand{\beqa}{\begin{eqnarray}}
\newcommand{\eeqa}{\end{eqnarray}}

\newcommand{\hinvMpc}{{h^{-1}{\rm Mpc}}}
\newcommand{\hMsun}{h^{-1}{\rm M_{\odot}}}

\newcommand{\hMpc}{h{\rm Mpc}^{-1}}
\newcommand{\hinvGpc}{{h^{-1}{\rm Gpc}}}

\begin{document}

\title{Measuring equality horizon with the zero-crossing of the galaxy correlation function}
\author{F.~Prada$^1$, A.~Klypin$^2$, G.~Yepes$^3$, S.~E.~Nuza$^4$, S.~Gottl$\ddot{\rm o}$ber$^4$}
\medskip
\affiliation{$^1$Instituto de Astrof\'{i}sica de Andaluc\'{i}a (CSIC), E-18080 Granada, Spain  \\ 
$^2$Astronomy Department, New Mexico State University, Las Cruces NM, USA  \\ 
$^3$Departamento de F\'{i}sica Te\'{o}rica, Universidad Aut\'{o}noma de Madrid, Spain  \\ 
$^4$Leibniz-Institut f$\ddot{\rm u}$r Astrophysik Potsdam (AIP), Germany}

\date{\today}

\begin{abstract}

The size of the horizon at the matter-radiation equality is a key scale of the Big Bang 
cosmology that is directly related to the energy-matter content of the Universe. In this letter, 
we argue that this scale can be accurately measured from the observed clustering of galaxies in new large scale surveys. We demonstrate that the zero-crossing, $r_c$, of the 2-point
galaxy correlation function is closely related to the horizon size at matter-radiation equality 
for a large variety of flat $\Lambda$CDM models. Using large-volume cosmological simulations, we also show that the pristine zero-crossing is unaltered by non-linear evolution of density fluctuations, redshift distortions and galaxy biases. This makes $r_c$ a very powerful standard ruler that can be accurately measured, at a percent level, in upcoming experiments that will collect redshifts of millions of galaxies and quasars.

\end{abstract}


\maketitle

\setcounter{footnote}{0}


The science exploitation of upcoming experiments such as BOSS, DES,
LSST, BigBOSS and Euclid that will survey the uncharted universe
taking spectra and images of millions of galaxies and quasars,
combined with the analysis of Grand Challenge cosmological
simulations and development of theoretical models, will be critical to 
accurately measure the properties of our Universe
(e.g. \cite{2008ARA&A..46..385F} for a
review).
\\
The physics governing the evolution of perturbations in CDM universes
imprints two distinct length scales in the galaxy distribution which
are useful as ``standard rulers'' for distance estimates: (1) the
scale of the particle horizon at matter-radiation equality and (2) the
sound horizon scale at the baryon-drag epoch before recombination
\cite{1998ApJ...496..605E,1999ApJ...511....5E}. The Baryonic Acoustic
Oscillations (BAO), related with the later, have been detected in the
SDSS, 2dFGRS and WiggleZ galaxy power spectra at a characteristic
scale $k_{\rm bao}>0.05\,\hMpc$, and as a single peak in the
galaxy two-point correlation function (CF) at $r_{\rm
  bao}\approx105\,\hinvMpc$ (e.g.     \cite{2005ApJ...633..560E,2005MNRAS.362..505C,2010MNRAS.401.2148P,
2009MNRAS.400.1643S,2011MNRAS.415.2892B}).
%

The total physical matter density 
($\omega_{\rm m}\equiv\Omega_{\rm m} h^2=\Omega_{\rm cdm}
h^2+\Omega_{\rm b} h^2$) and, thus, the moment of equality
can be measured from CMB fluctuations alone, but the
measurement errors decline almost by a factor of two if BAO
clustering estimates are included \cite{2011ApJS..192...18K}. It would be
helpful to have another method for testing the equality epoch, which
is more sensitive to  $\omega_{\rm m}$ than BAO.


The particle horizon at matter-radiation equality is
associated with the most prominent feature of the linear power
spectrum of density fluctuations  $P(k)$ - the turnover or maximum - at
the characteristic scale $k_{\rm max}$ $\sim 0.015 \,
\hMpc$. This corresponds to the transition from $P(k)\sim k$ 
for a Harrison-Zeldovich scale-invariant spectrum to a $P(k)\sim k^{-3}$
spectrum due to modes that were not growing after they entered the
horizon during the radiation dominated era
\cite{1980lssu.book.....P}. Thus, we expect that the location of this
turnover will be related to the scale of the horizon at
matter-radiation equality $k_{\rm eq}$
\cite{1999ApJ...511....5E}.

The equality wavenumber $k_{\rm eq}$ is given by $k_{\rm
  eq}=a_{\rm eq} \, H_{\rm eq}(a_{\rm eq})$ with $a_{\rm eq} =
(\omega_\gamma/\omega_{\rm m})(1+0.2271 N_{\rm eff})$ and $H_{\rm
  eq}(a_{\rm eq})=\sqrt{2\Omega_{\rm m}}H_{\rm 0}(1/a_{\rm eq})^{3/2}$
being the scale factor and expansion rate at matter-radiation equality
\cite{1998ApJ...496..605E}. We adopt a radiation density
$\omega_\gamma\equiv\Omega_\gamma h^2 = 2.469\times 10^{-5}$ and $
  N_{\rm eff} = 3.04$ for standard neutrino species
\cite{2011ApJS..192...18K}. Hence, it can be seen that the equality
horizon scale depends solely on the matter density as $k_{\rm
  eq}\propto\,\omega_{\rm m}$. The size of the horizon at the matter-radiation
equality epoch 
 is defined by the comoving distance $r_{\rm eq}\equiv D_{\rm H}=\int_0
   ^{ a_{\rm eq}}[a^{-2}/H(a)]da=(4-2\sqrt{2})\,k_{\rm
  eq}^{-1}$. Thus, $r_{\rm eq}$ scales with matter density as 
   $r_{\rm eq}\propto\,\omega_{\rm m}^{-1}$. We obtain $k_{\rm
  eq}\simeq0.014\,\hMpc$ and $r_{\rm eq}\simeq85 \, \hinvMpc$ adopting
the latest matter density observational constraints
\cite{2011ApJS..192...18K}.

\begin{figure}
\includegraphics[width=0.5\textwidth]{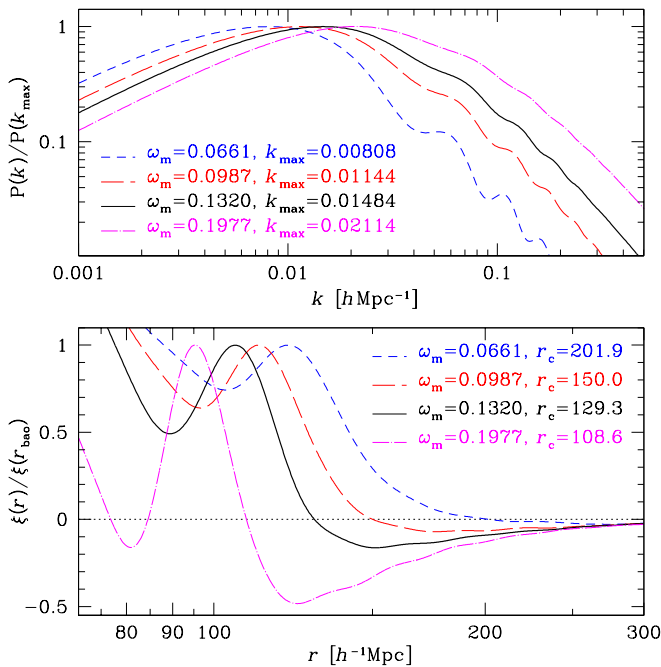}
\caption{\label{fig:fig1} Linear power spectra $P(k)$
  (top panel), and their corresponding correlation functions
   $\xi(r)$ (bottom panel) for  $\Lambda$CDM
  models with a fixed baryon density $\omega_{\rm
    b}=0.023$. Thick lines corresponds to our fiducial model.
   When $\omega_{\rm m}$ increases, the
  turnover in $P(k)$ and  its characteristic scale $k_{\rm
    max}$ moves to larger $k$. Thus, the zero-crossing in the
  correlation function $\xi(r_{\rm c})=0$, located beyond the
  BAO peak, shifts to smaller radii. For each model, both
  $P(k)$ and $\xi(r)$  have been normalized to its
  maximum and BAO peak respectively.}
\end{figure}

The turnover in $P(k)$ shifts to high wavenumbers for models with larger matter density mainly
as a result of the scaling relation between the equality wavenumber $k_{\rm eq}$ and $\omega_{\rm m}$. The matter perturbations remain frozen after entering the
horizon, but since this process is not instantaneous the transition
from the $P(k) \propto k$ to $P(k) \propto k^{-3}$ is broad over about an
order of magnitude in $k$.  For dark matter-baryonic models the expected scaling with 
matter density might involve an additional dependence on other cosmological parameters. 
In order to derive a relation between $k_{\rm max}$ at the maximum of $P(k)$ and the equality scale $k_{\rm eq}$, we computed the linear matter power spectrum of density
fluctuations of a flat $\Lambda$CDM universe, using CAMB
\cite{2000ApJ...538..473L}, for a number of
cosmological models with varying $\omega_{\rm m}$ and $\omega_{\rm b}$ 
around our fiducial model. For this fiducial model, which is consistent with WMAP7 
cosmological parameters and other recent cosmological constraints 
(\cite{2011ApJS..192...18K} and references therein), we assume the matter 
density $\omega_{\rm m}=0.132$, the
normalization of fluctuations $\sigma_{\rm 8}=0.82$, the primordial index of
scalar perturbations $n_{\rm s}=0.95$, the baryon density 
$\omega_{\rm b}=0.023$, the Hubble constant $H_0=100\, h \, {\rm km \,
  s^{-1}Mpc^{-1}}$ with $h=0.7$, and $N_{\rm
  eff}=3.04$. In Fig.~\ref{fig:fig1} (top panel), we show the linear matter power spectrum of our 
$\Lambda$CDM models, computed with CAMB, for different $\omega_{\rm m}$ values and 
kept fixed the other cosmological parameters that
determine $P(k)$, to that given by our fiducial model. As expected, the turnover location
$k_{\rm max}$ is close to $k_{\rm eq}$ and scales with $\omega_{\rm m}$. Yet, the 
precise turnover scale $k_{\rm max}$ depends slightly on other cosmological parameters 
such as the primordial tilt $n_{\rm s}$.

\begin{figure}[hb!]
\includegraphics[width=0.49\textwidth]{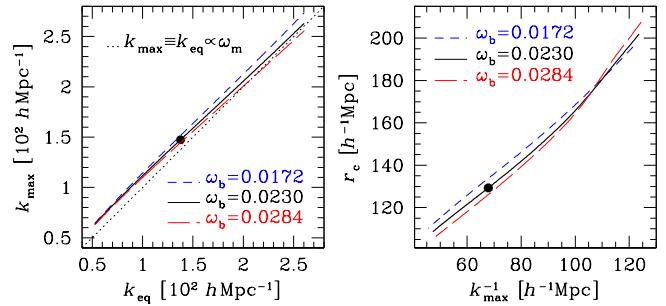}
\caption{\label{fig:fig2} $P(k)$ turnover location $k_{\rm max}$
  as a function of the horizon scale at matter-radiation equality
  $k_{\rm eq}$ (left panel), and its relation with the
  corresponding zero-crossing position $r_{\rm c}$ in the linear
  $\xi(r)$ (right panel) for $\Lambda$CDM models
  with three different baryon densities. Solid symbols are for
  our fiducial model.}
\end{figure}

We show in Fig.~\ref{fig:fig2} (left panel) the dependence of 
$k_{\rm max}$ as a function of the expected equality wavenumber
$k_{\rm eq}$ for different models with fixed baryon density
$25\%$ below and above that of our fiducial model
with $\omega_{\rm b}=0.023$ (solid line), i.e. $\sim6\sigma$ around
best observational constraints. Matter density $\omega_{\rm m}$ varies
over a large range from 0.05 to 0.25 around our fiducial value
$\omega_{\rm m}=0.132$ (solid symbol). 
 We kept fixed $\sigma_{\rm 8}$ and
$n_{\rm s}$. The dotted line represents the $k_{\rm max}\equiv k_{\rm
  eq}$ relation which scales only with $\omega_{\rm m}$ as $k_{\rm
  eq}=0.104\,\omega_{\rm m}\,\hMpc$. The difference between 
$k_{\rm max}$ and $k_{\rm eq}$ is less than $20\%$ for the range of parameters considered here. 
Models with smaller baryon content
show a better agreement between $k_{\rm max}$ and $k_{\rm eq}$.
The weak dependence of $k_{\rm max}(k_{\rm eq})$
on baryon density $\omega_{\rm b}$ can be approximated by
\begin{eqnarray}
 k_{\rm max}=  \left(\frac{0.194}{\omega_{\rm b}^{0.321}}\right)
      k_{\rm eq}^{0.685-0.121\,\log_{10}(\omega_{\rm b})}. \label{eq:eq1}
\end{eqnarray}
We also analyzed models with different primordial tilt $n_{\rm s}$ $\sim 2.5\sigma$ around the latest constraints \cite{2011ApJS..192...18K}, and found  only a small dependence of $k_{\rm max}$ on $n_{\rm s}$.

The existence of the turnover in the galaxy power spectrum has been
discussed extensively in the literature, although its detection has
not been reported up to now in any completed galaxy redshift survey
such as the 2dFGRS, SDSS or WiggleZ 
\cite{2005MNRAS.362..505C,2010MNRAS.404...60R,2011MNRAS.415.2892B}. Yet,
little attention has been paid to the imprint of the horizon scale at
matter-radiation equality on the zero-crossing in the galaxy two-point
correlation function $\xi(r)$. Although the very small clustering signal
at those large scales is severely affected by potential sources of
systematic errors 
\cite{2005MNRAS.363.1329B,2011MNRAS.tmp.1393R}, here, in this letter, we highlight 
that the zero-crossing deserves momentous attention both
from theory and observations. In the early 90s, Klypin \& Rhee
\cite{1994ApJ...428..399K} proposed to use the zero-crossing
measurements of the galaxy cluster CF as a
sensitive test for the shape of the power spectrum of initial
fluctuations. See also \cite{1997MNRAS.289..813E} for a detailed discussion on how, in their case, a sharp maximum in the power spectrum would be mapped 
into the correlation function.  

The detection of the zero-crossing in $\xi(r)$ is a
fundamental prediction of $\Lambda$CDM or any other dark matter-baryon
cosmological model \cite{1980lssu.book.....P,2002PhRvD..65h3523G,2004ApJ...615..573M}. 
The correlation function is related to $P(k)$ 
by the Fourier transform 
\cite{1980lssu.book.....P}:
\begin{equation}
  \xi(r) = \frac{1}{ 2\pi^2} \int_0^\infty dk k^2 P(k) \frac{\sin(kr)}{kr}.
\label{eq:eq2} 
\end{equation}
The correlation function must have a zero-crossing $\xi(r_{\rm c})=0$
at some radius because $P(k)\rightarrow0$ when $k\rightarrow 0$,
implying that $\int_{0}^\infty \xi(r)r^2 dr = 0$. Note that for our fiducial model, the
zero-crossing  $r_{\rm c}$ happens only at a scale
$\sim20\%$ larger than the position of the BAO peak, i.e. at 129.3 $\hinvMpc$  (see
Fig.~\ref{fig:fig1}, bottom panel).

\begin{figure}
\includegraphics[width=0.49\textwidth]{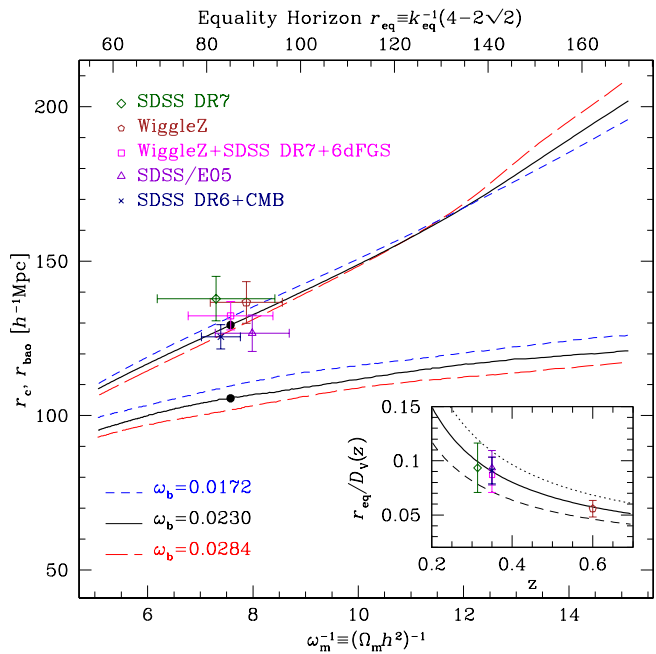}
\caption{\label{fig:fig3} Top curves show the zero-crossing scale
  ${\rm r}_{\rm c}$ as a function of matter density $\omega_{\rm
    m}^{-1}$ and equality horizon size $r_{\rm eq}$ (top axis)
  for our $\Lambda$CDM CAMB models with fixed values of $\omega_{\rm
    b}$. This relation can be well represented by a broken power-law
  as given in Eq.~(\ref{eq:eq3}).  As a reference, we also show the
  acoustic peak position $r_{\rm bao}$ (bottom curves). Solid symbol
  corresponds to our fiducial model. Zero-crossing estimates and
  1-$\sigma$ errors obtained from best-fit 
   $\omega_{\rm m}^{-1}$ and $\alpha$ stretch parameters from different 
    survey CF data are shown (open symbols). The
  inset panel shows the corresponding observational $r_{\rm eq}/D_V(z)$ estimates for
  different redshifts. Also displayed are predictions for $\Lambda$CDM
  models with $\omega_{\rm m}$ =0.11 (dashed line), 0.132 (fiducial,
  thick solid line) and 0.17 (dotted line).}
\end{figure}

However, the linear $\xi(r)$ for models with $\omega_{\rm m}\gtrsim0.2$ shows a
first zero-crossing located at scales $r_{\rm c}\lesssim80\,\hinvMpc$, i.e.
below the BAO feature (see Fig.~\ref{fig:fig1}). We recall that the SDSS and WiggleZ galaxy
CFs display a positive clustering signal well above
$\sim80\,\hinvMpc$, at least up to the BAO peak located at $\sim105\,\hinvMpc$, which already
imposes a matter density upper limit of $\omega_{\rm m}\sim0.2$ (the stacked galaxy CF of these surveys gives a BAO
detection at $4.9\sigma$ \cite{2011arXiv1108.2635B}). Here,
we study the zero-crossing in $\xi(r)$ located beyond the BAO feature
for models with $50\%$ matter density variation around our fiducial
value, i.e. $0.066 < \omega_{\rm m} < 0.198$.

Fig.~\ref{fig:fig2} (right panel) shows the relation between the
zero-crossing position $r_{\rm c}$ in the linear regime and its
corresponding $P(k)$ turnover location at $k_{\rm max}$ for 
models with $0.066 < \omega_{\rm m} < 0.198$.
It is clearly seen in Fig.~\ref{fig:fig3} that $r_{\rm c}$ is
determined by the scale of the horizon at matter-radiation equality.
Note, that ${\rm r}_c$ shows a much steeper increase with $\omega_{\rm
  m}^{-1}$ as compared with the position of the BAO peak $r_{\rm
  bao}$. Moreover, the BAO peak position shows a larger dependence on
baryon density $\omega_{\rm b}$, e.g. about $7\%$ and $3.5\%$
variation in $r_{\rm bao}$ and $r_{\rm c}$ respectively at fiducial
$\omega_{\rm m}=0.132$ (solid point). The zero-crossing displays a small dependence
on $\omega_{\rm b}$ as can be seen in Eq.~\ref{eq:eq1}. For our grid
of CAMB models with $0.066 < \omega_{\rm m} < 0.198$ and
$0.0172<\omega_{\rm b}<0.0284$ we obtain the following relation
between the zero-crossing and the comoving size of the
matter-radiation equality horizon, which is well described by a broken
power-law, i.e.
\begin{equation} r_{\rm c}=A\,r_{\rm eq}^{b_1}
  \Bigg [1+ \bigg ( \frac{r_{\rm eq}}{r_{\rm eq,0}} \bigg
  )^d \Bigg ]^{\frac{b_2-b_1}{d}},
\label{eq:eq3} 
\end{equation}
where $r_{\rm eq,0}=129.5\,\hinvMpc$, $b_1=0.425$, 
$d=6.5$. Parameters $A=14.478\,\omega_b^{-0.0785}$ , and
$b_2=34.456\,\omega_{\rm b}+0.171$ encode a slight
dependence on baryon density. The size of the equality horizon
 $r_{\rm eq}$ depends only on the equality characteristic scale
$k_{\rm eq}$ through the matter density, i.e. $r_{\rm
  eq}\equiv(4-2\sqrt{2})\,k_{\rm eq}^{-1}=11.231\,\omega_{\rm
  m}^{-1}\, \hinvMpc$.

To estimate the value of $r_{\rm c}$ with current galaxy surveys we use the constraints on the stretch parameter
$\alpha(z_{\rm eff})=D_V(z_{\rm eff})/D_{\rm V,fid}(z_{\rm eff})$ ($D_V(z)\equiv [(1+z)^2 D^2_A cz/H(z)]^{1/3}$ is the dilation distance \cite{2005ApJ...633..560E} and $D_A$ is the angular diameter distance), obtained by different groups from their
best-fit cosmology modeling to the observed SDSS and WiggleZ
redshift-space $\xi(s)$. Typically, this model fit is done on the redshift-scale range
$20\,\hinvMpc \lesssim  s  \lesssim 180\,\hinvMpc$.  This allows us to estimate the
expected position of the zero-crossing  using the relation
$r_c=\alpha(z_{\rm eff})\,r_{\rm c,fid}$. $D_{\rm
  V,fid}(z)$ and $r_{\rm c,fid}$ are the dilation distance at
$z_{\rm eff}$ (survey mean redshift distribution) and the zero-crossing scale for the fiducial cosmology.
Fig.~\ref{fig:fig3} shows our estimates taking constraints on
$\alpha(z_{\rm eff})$ and $\omega_{\rm m}$ from SDSS-LRG E05 \cite{2005ApJ...633..560E},
SDSS-LRG DR7 \cite{2010ApJ...710.1444K,2011arXiv1108.2635B},
WiggleZ $\&$ Stacked WiggleZ+SDSS+6dFGS
\cite{2011arXiv1108.2635B}; and SDSS-LRG DR6 combined with CMB \cite{2009MNRAS.400.1643S}.

\begin{figure}
\includegraphics[width=0.5\textwidth]{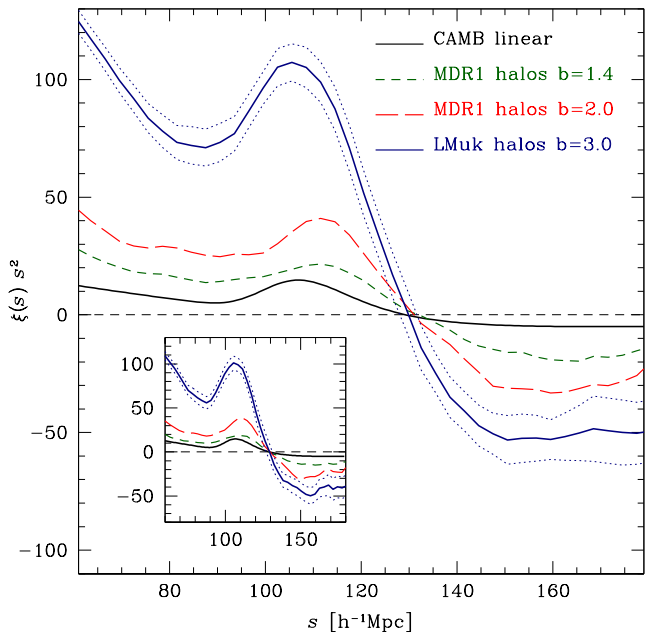}
\caption{\label{fig:fig4} Redshift-space correlation functions at
  $z=0.5$ for three halo (``galaxy'') samples with different bias in
  the MultiDark and LittleMuk $\Lambda$CDM simulations. Results are
  compared with the linear $\xi(r)$ (thick solid line) that corresponds
  to our fiducial cosmological parameters. The inset panel shows the
  real-space correlation function. The zero-crossing in real space is
  the same regardless of galaxy bias. There is a small $\sim 0.4\%$
  shift in the redshift-space. Dotted lines are statistical errors 
  estimated for "galaxies" in the volume of $46.875\,(\hinvGpc)^3$.}
\end{figure}

The accurate modeling and interpretation of both $P(k)$ and $\xi(r)$
galaxy clustering statistics present some difficulties. The shape of
the linear matter power spectrum $P(k)$ is distorted by the
nonlinear evolution of density fluctuations, redshift distortions and
galaxy bias even at large-scales $k < 0.2 \, \hMpc$
\cite{2003MNRAS.341.1311S,2006PhRvD..73f3519C,2008ApJ...686...13S,2008MNRAS.383..755A}. In
this regard we should also draw attention to the impact of those
inevitable effects on the expected zero-crossing scale in the
redshift-space galaxy correlation function, $\xi(s)$, as imprinted by
the horizon scale at matter-radiation equality.

We show in Fig.~\ref{fig:fig4} redshift-space
$\xi(s)$ at $z=0.5$ (close to the mean redshift of the BOSS and
WiggleZ galaxy distributions
\cite{2011MNRAS.415.2892B,2011ApJ...728..126W}), drawn from our
MultiDark and LittleMuk suite of $N$-body simulations of a flat
$\Lambda$CDM cosmology with the same parameters as for our fiducial cosmological
model. The MultiDark simulation (MDR1) was done using the ART code. It
has $2048^3$ particles in a $1\,\hinvGpc$ box (see
\cite{2011arXiv1104.5130P} and $\it{www.multidark.org}$ for
details). The mass and force resolutions are $8.72\times10^9\,\hMsun$
and $7\,h^{-1}{\rm kpc}$. LittleMuk (LMuk) consists of three GADGET \cite{2005MNRAS.364.1105S}
realizations of $1280^3$ particles of a larger box with
$2.5\,\hinvGpc$ on a side. Force resolution and particle mass are
$28\,{\it h}^{-1}{\rm kpc}$ and $5.58\times10^{11}\,\hMsun$. 
The LMuk total volume $46.875\,(\hinvGpc)^3$ allows us
to compute the mean and the variance for the estimates of $\xi(r)$ for
dark matter halos.  Dark matter halos (and subhalos) were
identified with the Bound-Density-Maxima (BDM) algorithm
\cite{2011arXiv1104.5130P}. We then use a simple, nonparametric abundance-matching
prescription, to connect dark matter halos (and subhalos) in our simulations with galaxies by selecting them above a given maximum circular velocity $V_{\rm max}$. This procedure is able to predict the clustering properties, and
therefore the two-point CF and halo occupation
distribution of observed galaxies for different number densities
(e.g. \cite{2006ApJ...647..201C,2010arXiv1005.1289T}). $V_{\rm max}$
thresholds
for the three ``galaxy'' samples with increasing biases indicated in the
plot are 180, 350 and 600 ${\rm km\, s^{-1}}$ roughly corresponding to
the Emission Line Galaxies (ELGs, $b=1.4$), Luminous Red Galaxies
(LRGs, $b=2.0$) and QSOs ($b=3.0$) in the major experiments discussed here.

Results presented in Fig.~\ref{fig:fig4} evidence that the
zero-crossing position $r_{\rm c}$ in the dark matter halo
CF appears at the location predicted by linear
theory.  Thus, $r_{\rm c}$ is not distorted by the non-linear
evolution of density fluctuations, redshift distortions and bias. This
makes the zero-crossing an invaluable tool to constrain the size of
the horizon at matter-radiation equality and the shape of the
primordial power spectrum of initial fluctuations. We measure
$r_c=129.8^{+7.2}_{-5.9}\,\hinvMpc$ from the mean LMuk ``galaxy''
sample (thin solid line in Fig.~\ref{fig:fig4}) with errors scaled
according to  statistics similar to that of the SDSS LRGs
($\sim10^5$ galaxies).
 Zero-crossing uncertainty goes down to a small value of
$\sim 1.5\%$ if we consider a survey with 10 times more galaxy statistics
such as BOSS (dotted lines) or $\sim 0.3\%$ in the case of 20 million galaxies, as
it will be targeted by the upcoming new surveys like BigBOSS or
Euclid.
Our estimates are consistent with Fig. 6 in
\cite{2008MNRAS.390.1470S} for  real- and redshift-space dark
matter and halo samples of similar bias 
if one compares  the expected zero-crossing 
$r_{\rm c}$ position for their adopted cosmology. The level of systematics should 
be small enough to allow accurate
measurements of $r_{\rm c}$.  We estimate that the error in $\xi(r)$
should be smaller than $\sim 10^{-3}(3\times10^{-4})$ for 4(1)\% error in
$r_{\rm c}$.

Several groups have reported the absence of negative signal,
and hence zero-crossing up to $\sim250\,\hinvMpc$ (see
\cite{2010ApJ...710.1444K,2009ApJ...696L..93M} and references
therein). It was speculated in \cite{2009A&A...505..981S}  that the
large-scale SDSS $\xi(s)$ is affected by intrinsic errors and
volume-dependent systematic effects. The implications of a possible
constant systematic shift in the redshift-space clustering 
due to unknown observational systematic errors
has been pointed out in
\cite{2009MNRAS.400.1643S}. 
It was argued in \cite{2010ApJ...710.1444K} that lack of zero crossing
in SDSS was simply due to cosmic variance, not unknown systematics.
Interestingly, the WiggleZ redshift-space $\xi(s)$ at $z=0.6$
\cite{2011MNRAS.415.2892B}, albeit with large uncertainties, displays a crossover around
$\sim120\,\hinvMpc$ and negative signal
above this scale up to $180\,\hinvMpc$, as expected for current matter
density observational constraints.

We believe that the detection of the zero-crossing in the galaxy
correlation function may offer some advantages as compared with 
finding the maximum in $P(k)$, though mathematically both are
equivalent.
The individual Fourier modes in $P(k)$ split the clustering signal in
fragments and as a result the signal-to-noise and accuracy are expected to be smaller as
compared with the cross-correlation signal around the zero-crossing
which is coming from the convolution of the entire $P(k)$ without any
binning or splitting of the signal. 

To summarize, adding zero-crossing estimates will improve the accuracy of
$\omega_{\rm m}$  because
$r_{\rm c}$ is more sensitive to $\omega_{\rm m}$ than the BAO peak
and because it does not require the elaborate reconstruction methods
used for BAO analysis \cite{2007ApJ...664..675E,2009PhRvD..79f3523P}. 

We thank A. Kravtsov, C. Blake, A. Sanchez, and D. Eisenstein for
fruitful discussions and comments. We are grateful to S. Knollmann for running LMuk initial conditions. MDR1
and LMuk simulations were run on NASA's Pleiades and
SuperMUC Migration system at LRZ M$\ddot{\rm u}$nchen. This work is supported by
Spanish (AYA10-21231, Multidark-CSD2009-0064, SyeC-CSD2007-0050, AYA09-13875),  German (DAAD, DFG-MU1020-16-1) and US NSF grants to NMSU.

\bibliographystyle{apsrev4-1}
\bibliography{draftzerop_v6}
\end{document}